\documentstyle[draft]{mn}

\input epsf

\begin{document}

\newcommand{\kms}{\,{\rm km}\,{\rm s}^{-1}}
\newcommand{\mum}{$\,\mu$m}
\newcommand{\muv}{$\,\mu$V}
\newcommand{\muk}{$\,\mu$K}
\newcommand{\arcm}{\,arcmin}
\newcommand{\arcs}{\,arcsec}
\def\gs{\mathrel{\raise0.35ex\hbox{$\scriptstyle >$}\kern-0.6em
\lower0.40ex\hbox{{$\scriptstyle \sim$}}}}
\def\ls{\mathrel{\raise0.35ex\hbox{$\scriptstyle <$}\kern-0.6em
\lower0.40ex\hbox{{$\scriptstyle \sim$}}}}
\def\ion#1#2{#1$\,${\small\rm{#2}}\relax}
\hyphenation{an-i-so-tro-py}
\hyphenation{an-i-so-tro-pies}
\hyphenation{an-i-sot-ro-pic}
\hyphenation{fluc-tu-a-tion}
\hyphenation{fluc-tu-a-tions}

\journal{Preprint UBC--COS--98--01, astro-ph/9808031}
\title[Using SCUBA to place upper limits on arcsecond scale CMB anisotropies
at 850\mum]
{Using SCUBA to place upper limits on arcsecond scale CMB anisotropies at
850\mum}

\author[C.\ Borys et al.]
       {Colin Borys, Scott C.\ Chapman and Douglas Scott
        \vspace*{1mm}\\
        Department\ of Physics \& Astronomy,
        University of British Columbia,
	6224 Agricultural Road,
        Vancouver, B.C.~V6T 1Z1,~~Canada\\
        }

\date{Accepted ... ;
      Received ... ;
      in original form ...}
\pubyear{1998}
\pagerange{000--000}
\maketitle

\begin{abstract}
The SCUBA instrument on the James Clerk Maxwell Telescope
has already had an impact on cosmology by detecting relatively large
numbers of dusty galaxies at high redshift.  Apart from identifying
well-detected sources, such data can also be mined for information about
fainter sources and their correlations, as revealed through low level
fluctuations in SCUBA maps.  As a first step in this direction we analyse
a small SCUBA data-set as if it were obtained from a Cosmic Microwave
Background (CMB) differencing experiment.  This enables us to
place limits on CMB anisotropy at 850\mum.
Expressed as $Q_{\rm flat}$, the quadrupole expectation value for a flat
power spectrum, the limit is 152\muk\ at 95 per cent confidence, corresponding
to $C_0^{1/2}{<}\,$355\muk\ for a Gaussian autocorrelation function,
with a coherence angle of about 20--25 arcsec; These results could
easily be reinterpretted in terms of any other fluctuating sky signal.
This is currently the best limit for these scales at high frequency, and
comparable to limits at similar angular scales in the radio.
Even with such a modest data-set, it is possible to put a constraint on
the slope of the SCUBA counts at the faint end, since even randomly
distributed sources would lead to fluctuations.
Future analysis of
sky correlations in more extensive data-sets ought to yield detections,
and hence additional information on source counts and clustering.
\end{abstract}



\section{Introduction}
\label{sec:intro}
Since COBE discovered the existence of anisotropy in the Cosmic Microwave 
Background Radiation (CMB) at large angular scales, balloon and ground
based experiments have detected anisotropy at a range of smaller scales (see
e.g.~Smoot \& Scott~1998).
All cosmological models predict that the very smallest scales should be free
of primordial anisotropy, because photons in small-scale
overdensities which entered the horizon early have been able to random walk 
out of the potential wells, and also because fluctuations on scales below
the thickness of the last scattering surface are suppressed.

Nevertheless, secondary anisotropies can be generated through a wide variety
of physical processes occurring at redshifts $<1000$.  For an overview of
some of these effects see Bond (1996).  There is new motivation from the
detection of the Far Infrared Background (FIB, Puget et al.~1996,
Schlegel, Finkbeiner \& Davis~1998, Fixsen et al.~1998, Hauser et al.~1998),
as well as recent SCUBA results, that
dusty galaxies at high redshifts could be of greater significance than
previously assumed -- possibly the dominant source of CMB anisotropy at
arcsecond scales, and certainly the dominant source of sky fluctuations at
$\sim1\,$mm, at least out of the galactic plane.  The smooth FIB is
expected to break-up into sources, plus fluctuations due to unresolved sources,
and correlations between sources on the sky.  Weak Sunyaev--Zel'dovich
increments, caused by hot gas in virialised structures, may also contribute to
anisotropies in the sub-millimetre, and of course other physical effects
could also be important.
We attempt here to use SCUBA data to place limits on general fluctuations
within the framework of CMB anisotropies, i.e.~as intensity fluctuations on a
$2.728\,$K blackbody spectrum; but everything can be reinterpretted in terms
of fluctuations of some other spectral component.

Experiments probing similar angular
scales have already placed upper limits of anisotropy
at a variety of wavelengths, and a summary can be found in
Table~1.
The results are generally quoted in terms of the most
sensitive GACF (which will be described in Section~\ref{sec:likely}),
although a few experiments simply give a limit to the variance of the
temperature measurements.  Limits set at radio wavelengths are markedly more
sensitive than those in the millimetre region.  However, the expected
foreground signals at low frequency are expected to be entirely
different to those which are likely to dominate fluctuations in the
sub-millimetre sky (namely dusty galaxies and hot cluster gas).
The most relevant previous measurement is the
earlier JCMT limit of Church, Lasenby \& Hills (1993), using the single
bolometer which was the predecessor to SCUBA.

%
%
\begin{table*}
\label{tab:otherwork}
\begin{center}
\caption{ \hfil Summary of previous small angular scale CMB limits.\hfil }
\begin{tabular}{rrrll}
\noalign{\medskip}
\noalign{\smallskip}
$\lambda$  & Angular Scale & ${\Delta T}/T$ & Instrument & Reference\cr
\hline
800\mum\    &  17\arcs\   & $\leq 146\times10^{-5}$      & JCMT  &
 Church et al.~(1993)\cr
1250\mum\   &  30\arcs\ & $\leq 14\times10^{-5}$       & SEST \& IRAM  &
 Andreani (1994)\cr
           &  70\arcs\ & $\leq 24\times10^{-5}$       &       & \cr
           & 140\arcs\ & $\leq 19\times10^{-5}$       &       & \cr 
1300\mum\   &  12\arcs\ & $\leq 18\times10^{-5}$       & IRAM  &
 Kreysa \& Chini (1989)\cr
2110\mum\   &  66\arcs\ & $\leq 2.1\times10^{-5}$      & SuZIE &
 Church et al.~(1997)\cr
3400\mum\   &  10\arcs\ & $\leq 9\times10^{-5}$        & IRAM  &
 Radford (1993)\cr
15{,}000\mum\      & 156\arcs\ & $\leq 1.7\times10^{-5}$      & OVRO  &
 Readhead et al.~(1989)\cr
19{,}700\mum\      & 20\arcs\  & $\leq 2.3\times10^{-5}$      & Ryle  &
 Jones (1997)\cr
35{,}000\mum\      &  30\arcs\ & $\leq 2.3\times10^{-5}$       & ATCA  &
 Subrahmanyan et al.~(1998)\cr
           &  60\arcs\  & $\leq 1.6\times10^{-5}$      &       & \cr
           &  120\arcs\ & $\leq 2.5\times10^{-5}$      &       & \cr
35{,}000\mum\      &   6\arcs\ & $12.8\times10^{-5}$       & VLA   &
 Partridge et al.~(1997)\cr
           &  18\arcs\ & $\leq 4.8\times10^{-5}$   &       & \cr
           &  60\arcs\ & $\leq 2.0\times10^{-5}$   &       & \cr
\hline
\end{tabular}
\end{center}
\end{table*}
\bigskip

\section{SCUBA Observations}
\label{sec:scuba_obs}
The observations were conducted with the Submillimeter Common-User
Bolometer Array \cite{Cunnetal,GeaCun,Ligetal}
on the James Clerk Maxwell
Telescope.  SCUBA contains a number of detectors and
detector arrays cooled to 0.1 K that cover the atmospheric windows
from 350\mum\ to 2000\mum.  

\begin{figure}
\begin{center}
\leavevmode \epsfysize=8cm \epsfbox{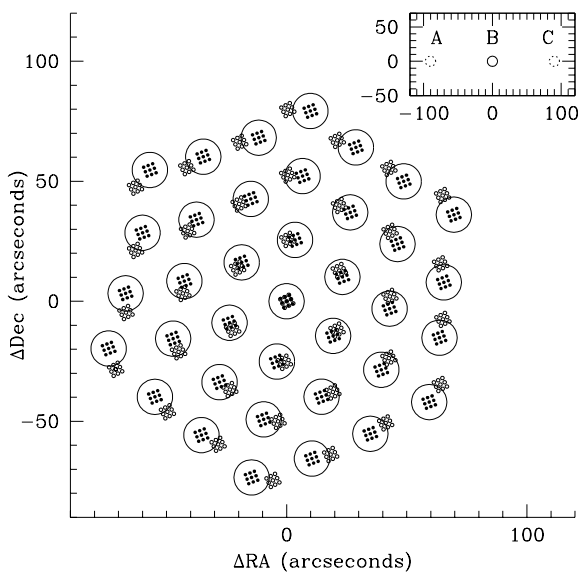}
\end{center}
\caption{The arrangement of the 37 bolometers in the long wavelength array.
The squares
of 9 points are the jiggle positions that create an effective beam of 
14.7\arcs, which are shown as the larger circles on the main plot.  The solid
jiggle pattern shows the bolometer positions for the first integration.  Due
to earth rotation, the entire array rotates throughout the course of the
run, and we have plotted the position of the bolometers for the last 
integration using open circles.  The inset illustrates the three positions
for the three beam chop that constitutes a single integration.  The entire
array is chopped in azimuth with this double-difference pattern.}
\label{fig:array}
\end{figure}

The arrays have a hexagonal arrangement of pixels, with feeds about two 
beamwidths apart in the focal plane.
The two array detectors provide an instantaneous field-of-view of 
2.3\arcm\ and can be used simultaneously. They consist of the 91  element 
Short-wave array, which we used with the 450\mum\ filter, and the 37
element Long-wave array, which we used at 850\mum. An illustration of the
observing strategy is given in Fig.~\ref{fig:array}, and described in more
detail below.  Essentially it is a double difference experiment, with some
additional complications.

During an integration, the telescope is `jiggled' in a 3 $\times$ 3 square 
pattern (with 2 arcsec offsets, shown in Fig.~\ref{fig:array}).
This improves the photometric accuracy, 
reducing the impact of pointing errors by averaging the source signal over an 
area slightly larger than the beam.  The effective half-power beamwidths 
for the 450 and 850\mum\ pixels are then 7.5\arcs\ and 14.7\arcs\ respectively.

Whilst jiggling, the secondary was chopped at 7.8125\,Hz between two 
positions, $A$ and $B$, separated here by 90\arcs\ in azimuth, yielding the 
single difference measurement $I_B-I_A$.  This is followed by a second single
difference, $I_C-I_B$, after the nod to the other side of $B$.
By differencing these two differences, we obtain a series of 
double differences, with the signal defined by $S=-I_A+2I_B-I_C$.  
For some of the systematic checks we will consider the data on a jiggle by
jiggle basis.  However, the data are generally treated by lumping the 9
jiggles together, yielding an 18 second double difference for each bolometer,
which we refer to as a single `integration'.  Considered in this way the
set-up is the traditional triple beam arrangement common in CMB
experiments.

SCUBA can also use a 64 point jiggle in order to fill in the blank space 
between the pixels (`mapping mode'),
but for this initial study, we use `photometry'
observations at 450 and 850\mum. Although this provides an undersampled
map at both wavelengths, it is more straightforward to analyze.

On 1997 December 3, five `scans' of 900 seconds, consisting of 50 
integrations each were taken.
The central pixel of SCUBA was fixed on a point source in an otherwise 
blank field, for which we were interested in obtaining sub-mm photometry 
(the lensed AGN B1933+503, Chapman, Fahlman \& Scott, in preparation).
The other pixels rotated relative to the sky through the period of the
integration; we show the positions of the first and last integrations
in Fig.~\ref{fig:array}.  Pointing was checked
hourly on the blazar 2036--419 and a sky-dip was performed between each 
15 minute scan to measure the atmospheric opacity.  
The rms pointing errors were below 2 arcsec, while the average 
atmospheric zenith opacities at 450\mum\ and
850\mum\ were fairly stable with $\tau$ being 0.51 and 0.12 respectively.
However, there were some short time-scale variations, presumably due to
water vapour pockets blowing over at high altitude, which caused some
parts of the data-set to be noisier (see Fig.~\ref{fig:rawtime}.
The observations were largely reduced using the Starlink package SURF 
(Scuba User Reduction Facility, Jenness \& Lightfoot~1998).
Spikes were first carefully rejected from the double difference data. The 
data were then corrected for atmospheric opacity and calibrated against 
Saturn and the compact \ion{H}{II}
region K3--50, which were also observed during 
the same observing shift. The  850\mum\ calibrations agreed with each 
other and also the standard gains to within 10 per cent. However, at 450\mum, 
K3--50 is extended and variable, and is not a good calibration source,
while the Saturn 450\mum\ calibration agreed with standard gains to within
25 per cent.

\section{Data Reduction}
\label{sec:data_reduc}
It is necessary to identify and remove any non-astronomical signals from the
data.  One of SCUBA's many strengths is its ability to redundantly
measure the strongest
of these contaminants: atmospheric emission.  To a lesser extent, cosmic ray
hits influence the data, but these are readily identified by the spikes they
leave in the timestream, and are therefore easily removed from the data (more
details in the next section).  Finally, we 
test for other correlations in the data that indicate a common signal, such
as cross-talk between bolometers.

In order to discuss our analysis procedure, we adopt the following notation.
The indices $b,s,i,$ and $j$ denote
bolometer, scan, integration, and jiggle number respectively, each starting at
1.  The variable $N$ subscripted with one of these indices represents the total
number of the quantity.  For this particular data-set they are:
$N_b$\,=\,37 or 91 (for the 450\mum\ and 850\mum\
channels respectively); $N_s$\,=\,5; $N_i$\,=\,50; and $N_j$\,=\,9.
It will also
be useful to introduce the variable $k=(s-1)N_i + i$, which simply indexes a
particular integration.

\subsection{Removing the atmospheric signal}
\begin{figure}
\begin{center}
\leavevmode \epsfysize=8cm \epsfbox{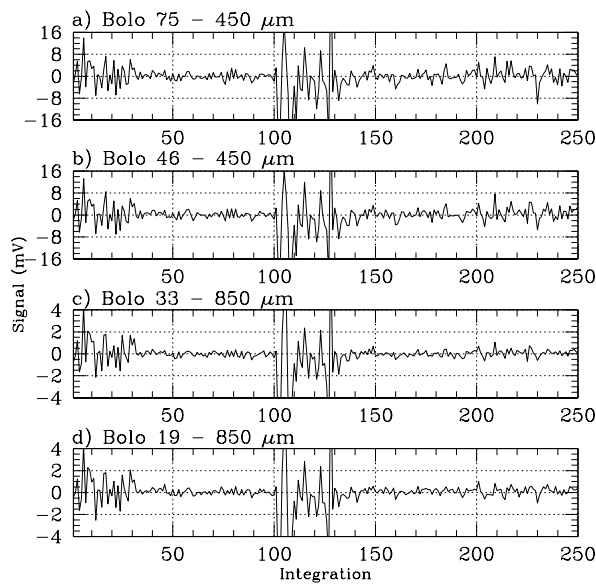}
\end{center}
\caption{The raw timestreams from 2 bolometers on each of the 450\mum\
and 850\mum\ arrays.
The central pixels at 450\mum\ and 850\mum\ are shown in panels b) and d).
The pixels plotted in panels a) and c) were chosen to be far from the
central pixel to help illustrate the correlated signal caused by the
atmosphere.  Vertical dashed lines separate the 5 `scans' (sets of 50
integrations).}
\label{fig:rawtime}
\end{figure}

We plot the signal timestream for two representative bolometers from each of
the short and long-wavelength array in Fig.~\ref{fig:rawtime}.  It is
clear that the output is highly correlated in time between bolometers,
even at different wavelengths.  Furthermore,
a detailed inspection of all the timestreams shows that the correlation
is strong even for bolometers at opposite
ends of the array, indicating that a common atmospheric
signal subtends an angle greater
than 2 arcmin.  Order of magnitude considerations suggest that the size
of a relevant patch is perhaps 1000 arcsec (e.g.~Jenness, Lightfoot
\& Holland~1998).
Much of this signal is removed in the process of chopping and nodding, but
some atmospheric noise inevitably remains (e.g.~Duncan et al.~1995).
It is reasonable to use the correlation across the array to calculate
a common atmospheric signal that can be removed from the data.
We will now consider two separate methods of removing the atmospheric signal
from the long wavelength data,
namely use of the average across the long wavelength array, or of the
independent average from the short wavelength array.

\subsubsection{Using the 850\mum\ data to remove atmospheric signals}
\label{sec:mean850}
The raw data contains
the 2 second double difference signal, $v_{bkj}$ which we use to
compute a mean sky signal on a jiggle by jiggle basis:
\begin{equation}
\label{equ:w_mean}
M_{kj}= {{1}\over{N_b-N_b^E}}{\sum_{b=1, b\not\subset E}^{N_b} v_{bkj}}.
\end{equation}  
Here
$N_b^E$ is the number of bolometers in the set $E$, which is a list of
all bolometers that are excluded from this sky calculation.  These include the
central pixel (which is `contaminated' by B1933+503) and any bolometers
that exhibit excessive noise.  These were chosen by looking at the
sky-corrected, integrated signal where the mean was computed leaving out only
the central pixel.  Those excluded show an integrated variance about twice as
high as the other bolometers in their channel.  For the 850\mum\ data, 7
bolometers were
removed ($E{=}\left\{12,19,22,23,24,32,37\right\}$) and the integrated signal
changed by only 1\muv, with a comparable reduction in the error bar.
At 450\mum, $E{=}\left\{46,87,8,59,62,64,65,66,67,69,70,71,78,79\right\}$
and the integrated signal changed by no more than 4\muv.
The sky-corrected signal is then simply
\begin{equation}
\label{equ:signal}
V_{bkj}=v_{bkj}-M_{kj}.
\end{equation}

The 9 jiggles at each integration are now binned together by taking a direct
mean to form the signal $V_{bk}$. Because the readout noise is very stable for
a given scan, we can assume that the noise in each integration is drawn from
the same random distribution.  If we further assume that this noise is
uncorrelated from integration to integration, we can compute the mean and
variance of the sky-corrected signal for each scan and each bolometer using
the data, via the scatter of the points within a scan:
\begin{equation}
\label{equ:Vbar}
{\overline V}_{bs}= {{1}\over {N_i-N_{bs}^E}}{\sum_{i=1}^{N_i} V_{bsi}q_{bsi}}
\end{equation}  
\begin{equation}
\label{equ:sigmaV}
{\rm and}\qquad {(\sigma_{bs}^V)}^2=  {1\over {N_i-N_i^E-1}} \sum_{i=1}^{N_i}
\left(V_{bsi} - {\overline V}_{bs} \right)^2 q_{bsi}.
\end{equation}  
Integrations $k=1$--40 and $k=101$--140 (the first parts of scans 1 and 3) were
taken while clouds were obscuring the target region
(see Fig.~\ref{fig:rawtime}), rendering them unusable
because the rapidly changing opacity cannot be easily characterised.  To
account for
these data, we introduce the quantity $N_{bs}^E$ which is the total number of
bad integrations in a particular scan for a given bolometer, and the
quantity $q_{bsi}$, which  takes
the values 1 and 0 for good and bad data respectively.   For our particular
data set, there are then
$80 \times N_b=$ 2960(7280) integrations at 850(450)\mum\ that are not used in
the subsequent analysis. 

\begin{figure}
\begin{center}
\leavevmode \epsfysize=8cm \epsfbox{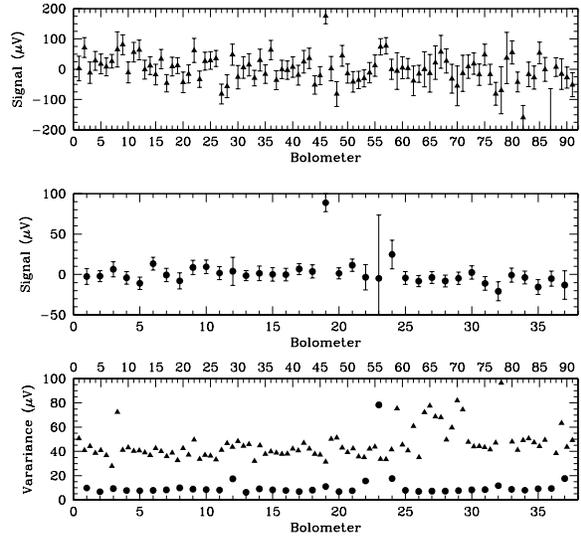}
\end{center}
\caption{The integrated signal measured by each bolometer on the 450\mum\ array
is shown in the top panel and the 850\mum\ array in the centre panel.
The detection in the central pixels, bolometer 19 at 850\mum\ and 46 at
450\mum, is readily apparent and due to the known source at that position.
The variance of each bolometer is also
plotted separately in the bottom panel, to clearly identify noisy bolometers.}
\label{fig:integrate}
\end{figure}

Anomalous signals (e.g.~cosmic ray hits) in the corrected
data were removed on a bolometer by bolometer, scan by scan basis.
Any integration that deviated by more than $3\sigma_{bs}^V$ was removed and
the variance recalculated.  This procedure was repeated once more to ensure that
any statistically significant anomalies shadowed by even larger ones were
removed during the second pass.  The total number of integrations removed was
31/2 (44/4) for the two passes at 850(450)\mum.  A third pass failed to remove
any additional points.  The removal of noisy sky sections and anomalous
signals left 68 per cent of the data to be used in subsequent analysis.
Finally, Fig.~\ref{fig:integrate} plots the integrated signal for each
bolometer, which is calculated using the weighted mean
\begin{equation}
\label{equ:int_mean}
I_b= {{\sum_{s=1}^{N_s}\sum_{i=1}^{N_i} V_{bsi}{(\sigma_{bs}^V)}^{-2}}\over
{N_i\sum_{s=1}^{N_s} {(\sigma_{bs}^V)}^{-2}}}.
\end{equation} 
Note that the likelihood analysis in Section~\ref{sec:likely}
does not simply use this binned version of the data, but allows for
full spatial correlations on the sky.

The method of removing atmospheric signal described above
is very successful, and some variant of it is the
usual method adopted for reducing SCUBA data.  One disadvantage however, is
that its use requires knowing in advance which pixels contain signal, or
using some iterative process to remove the sky when low-level signals are
present.  Another disadvantage, peculiar for our purposes here,
is that this method
correlates all of the pixels, since the mean has been subtracted from each one.
It is possible to take this into account in a full likelihood analysis of the
fluctuations by essentially also removing a mean from the theory.  However,
we simply ignored these correlations, realizing that they will be 
negligible for the bin sizes used in our analysis.  Each binned data point
uses $\sim N_bN_i$ integrations in determining the mean signal, introducing
a tiny correlation that is the inverse of this quantity.  Also, any 
correlations in the data will show up as a signal in the likelihood plots of 
Section~\ref{sec:likely}.
Since we ultimately find that the signal is consistent with zero,
we can safely assume 
that our mean  subtraction approach is sufficient.  Were this not the case, 
we would have to make the analysis insensitive to the mean using a matrix
rotation method \cite{bunnetal}, or marginalization \cite{BonJK}.  
 
\subsubsection{Using the 450\mum\ array as an atmospheric monitor}
We can avoid the correlation problem altogether by using an independent
estimate for removing the baseline
from the data.  The mean sky signals calculated
from equation~(\ref{equ:w_mean}) are extremely correlated for the 450\mum\
and 850\mum\ data, and therefore it is feasible to attempt
to subtract the sky
using the independent data from the other channel.  This may be particularly
useful for future SCUBA cosmology studies, since for `blank' fields
the 450\mum\ data will generally contain no
signal (while the 850\mum\ data may contain a contribution from extragalactic
sources).  Hence this method may have general utility for helping to look for
low levels of fluctuations in long integrations at 850\mum, where it becomes
important not to remove any of the signal along with the sky.

\begin{figure}
\begin{center}
\leavevmode \epsfysize=8cm \epsfbox{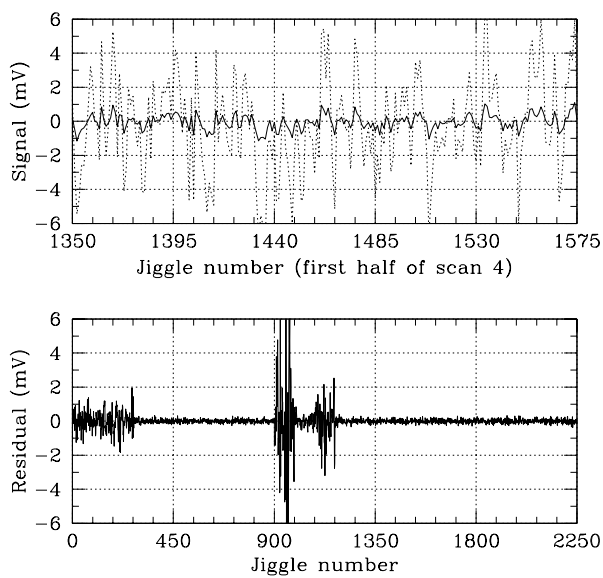}
\end{center}
\caption{Using the 450\mum\ array to remove the atmosphere from the 850\mum\
array.  The top panel shows the mean sky signals for both arrays over a subset
of the timestream, with the 850\mum\ data solid and the 450\mum\ data dashed.
The signals are clearly correlated.  The bottom panel
demonstrates the small residual between the 850\mum\ mean and scaled 450\mum\
mean.}
\label{fig:mean}
\end{figure}

The 450\mum\ channel is more susceptible to changes in opacity,
so we divide each scan into 10 sections with 5 integrations each,
and perform a least-squares fit of the form
\begin{equation}
\label{equ:chifit}
\chi^2= \sum_k\sum_j \left[M_{kj}^{850} -
 \left(cM_{kj}^{450} + o\right)\right]^2,
\end{equation}  
where $k$ runs over the 10 integrations to be fit (excluding, of course those
integrations
dominated by noisy atmospheric signal) and $j$ over each jiggle point in the
integration.  The value of $c$ is typically near 0.2 (see also Jenness
et al.~1998), and does not vary by more than 10 per cent within a scan.
The offset $o$ is on the order of a few \muv, and we found it necessary to
include it;
If we redo the fit and fix $o=0.0$, the integrated signal tends to be
biased lower by approximately 4\muv.
In Fig.~\ref{fig:mean} we plot the 450\mum\ and 850\mum\ mean sky for a section
of scan 4.  Also in the figure we plot the residual between the 850\mum\ mean
sky and the scaled 450\mum\ data.  To appreciate how small this residual
is, we take as an example the rms of the residual for the 450 jiggles in
scan 4, which turns out to be 100\muv.  The rms of the sky-corrected signal
for a typical bolometer in scan 4 using either the 850\mum\ or 450\mum\ mean is
about 380\muv.

\begin{figure}
\begin{center}
\leavevmode \epsfysize=8cm \epsfbox{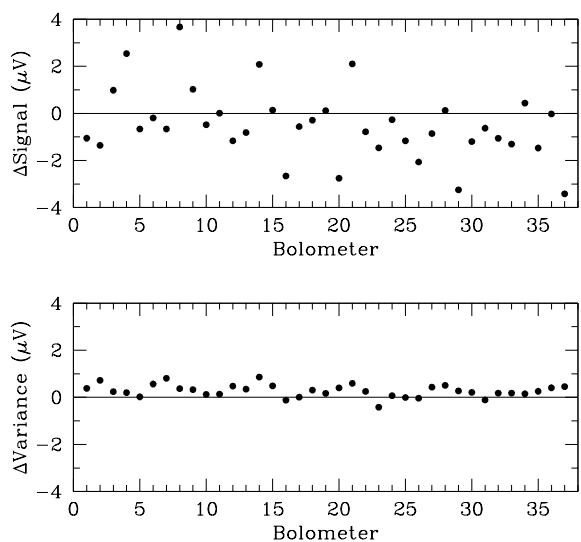}
\end{center}
\caption{The difference in the integrated signals between 850\mum\ data
sky-subtracted using the 850 and 450\mum\ array averages.}
\label{fig:resid}
\end{figure}

After forming the new mean signal, we subtract it from the 850\mum\ data and
calculate the integrated signal.  As evident in Fig.~\ref{fig:resid}, the
signal level changes by at most 4\muv\ compared to the results obtained by
using the 850\mum\ data to subtract a sky signal (top panel).
There does not appear to be
any bias introduced due to this method, although the variance is systematically
higher by about 0.5\muv\ or $0.12\,$mJy (bottom panel).
The increased variance is consistent with a signal 
with an additional independent noise term about one-third the size of the 
main noise term, as we can predict from the residual of the 850 and
450\mum\ mean signal level.

Therefore this method invariably introduces more uncertainty in the integrated
signal, but may be the best option when there is, for example,  extended 
structure in the 850\mum\ map, and nothing but noise in the 450\mum.  In this
case, however, care must be taken to avoid removing a DC level in the 850\mum\
data via the parameter $o$, which is a systematic effect that could be
calibrated by analyzing other data sets that are largely
free of sources in both
channels.  For our data set, this additional variance is only 5 per cent, and
influences our likelihood fits in Section~\ref{sec:likely}
by roughly this amount.

\subsubsection{Testing for a plane through the bolometer array}
It is plausible that the two dimensional bolometer array exhibits evidence of
a systematic temperature gradient, caused either by atmospheric signal 
or by sources even more local to the telescope.  This might introduce extra
variance into the data, which could erroneously be identified with larger
scale signal.
To clarify this, we attempt to fit the data for each array using 
\begin{equation}
\label{equ:fitplane}
P(x,y)=A + B x + C y
\end{equation} 
where $x$ and $y$ are the differences in altitude and azimuth 
respectively from the central beam.

Having found the best fitting plane at each time interval, we want to
assess whether it is real (i.e.~due to the atmosphere or some other
systematic effect) or just statistical (i.e.~due to stochastic noise).
The simplest approach is to test the
hypothesis that a for a given integration there exists a correlation 
between the best fit plane in the 450 and 850\mum\ data, as would be
expected if the plane is a genuine atmospheric effect.  To implement this,
we apply a Spearman rank correlation test to the following two statistics:

\noindent
the normalised height of the edge of array with y=0,
\begin{equation}
\label{equ:xedge}
X(\lambda)={{B\times70^{\prime\prime}}\over{A}},
\end{equation}

\noindent
and at x=0,
\begin{equation}
\label{equ:yedge}
Y(\lambda)={{C\times70^{\prime\prime}}\over{A}}.
\end{equation}
Here 70 arcsec is the approximate radius of the SCUBA array.  

The results are somewhat more complicated than we expected.
Performing the analysis for each integration,
we find a highly statistically significant correlation of $r=0.44$
between planes in the azimuthal direction, but with a relatively
small slope: 
(${\overline Y}(850\mu\rm m)=0.01$; ${\overline Y}(450\mu\rm m)=0.04$).
This small residual slope is not surprising, considering that a double
difference measurement inherently removes gradients, but the significance
of the correlation (formally $P=2\times10^{-9}$ was unexpected).

The correlation and significance parameters of the test performed on the
altitude direction are comparable ($r=0.33$, with a probability of
$P=1\times10^{-5}$), however, the slopes of the planes are, on
average, more pronounced: 
${\overline X}(850\mu\rm m)=-0.11$; ${\overline X}(450\mu\rm m)=0.01$.

These results certainly indicate that there is a residual slope across
the data, most likely as a result of atmospheric variation.
The fit for the 450\mum\ plane is generally better (as
verified by inspecting of the reduced chi-square statistic),
because there are more bolometers in the short-wavelength array.
Particularly at 850\mum\ the best-fitting
plane for a particular integration will
be largely a stochastic variation, although taking off these planes will
statistically remove some atmospheric signal.

We also looked at whether there were correlations in the planes when we
binned integrations together, in order to bring down the stochastic variation
in the planes.  We found that the highest correlation we 
could obtain was $r=0.60$ when the bin size was 5 integrations.  However, 
there are then only $170/5=34$ planes to use in the rank correlation test,
making the probability derived from the rank correlation questionable.  Indeed,
this fit did a poor job of cleaning up the signal, because the rms after
sky subtraction using this plane was twice as large as the single integration
plane.
With the data binned into 5 integrations we found the azimuthal correlation
between the 450 and 850\mum\ planes to be not very significant, while
the altitude slope correlation was still very highly significant.

The conclusion is that some evidence of residual sky-planes exists in our
data set, perhaps stronger in the altitude direction than the azimuth
direction.  These effects, however, are quite small, and are almost
negligible for our analysis of fluctuations.  Nevertheless we will regard
the data-set with sky-planes removed as providing our best estimate of any
residual fluctuations.  Clearly this issue merits further investigation with
more comprehensive SCUBA data-sets.

\subsection{Testing correlations between the bolometers}
Systematic effects might introduce correlations between bolometers
in the {\it time} domain, while genuine astronomical signal shows up as
correlations in the {\it space} domain.
There are three types of potential correlations introduced by the instrument
for which we would like to test: cross-talk between the bolometers 
grouped 16 per A/D card, crosstalk between exceptionally noisy bolometers
and their neighbours (listed in Section~\ref{sec:mean850}),
and possibly spill-over
from the central pixel which contains the signal from B1933+503.

The cross-correlation coefficient, $r_{XY}$ of the sky-corrected signal 
was taken  between each pair of data points. Here, we are using the 170 
18--second integrations calculated previously.  We find that $|r_{XY}|<0.3$ 
except for bolometers 32 and 36, which have a fairly strong positive
correlation of 0.5.  In fact these two bolometers are adjacent to each
other in the array {\it and} on the same A/D card.  In any case,
we find that removing data from this pair (or any other pair)
of bolometers affects the final results by no more than 20 per cent.  
However, we argue that the correlation is not strong enough to say with 
certainty that cross-talk is the cause.  It is of course possible that both
bolometers are observing an unresolved source with a flux slightly
below the noise level (we will return to this issue in
Section~\ref{sec:pointsource},
when we discuss the impact of unresolved point 
sources on the CMB fluctuations).
For that reason we do not exclude these bolometers from our analysis.

\subsection{Calibration and conversion of data into Thermodynamic temperature}
\label{sec:calibration}
Calibration was performed using sources of known flux during the same
observing period.  Specifically we used Saturn and the \ion{H}{II} region
K3--50.  Corrections were also made for the zenith opacity and the elevation
angle.  As discussed in Section~\ref{sec:scuba_obs}, these calibrations
agreed closely with the `standard gain' for the system at 850\mum.  At
450\mum\ the agreement was not so good, but the shorter wavelength data
were less important for our analysis in any case.  We list in
Table~2
the conversions between voltage and flux density which we used.

The measured intensity of a blackbody at a thermodynamic temperature,
$T_{\rm CMB}$, is given by
\begin{equation}
\label{equ:Ibb}
I_{\nu}={{2h\nu^3}\over{c^2}}{{1}\over{e^x-1}},
\end{equation} 
where $x=h\nu/kT_{\rm CMB}$.  To get the flux density in the beam, we multiply 
this by the solid angle of the beam: $S_{\nu}=I_{\nu}\Omega$.
Assuming that the beam is Gaussian, the solid angle is
$\Omega=(2\pi/8\ln2)\,{\rm FWHM}^2\simeq1.133\,{\rm FWHM}^2$.
Knowing the intensity at a given frequency $\nu$,
we can calculate the temperature of the source relative to the CMB by taking
the derivative of equation~(\ref{equ:Ibb}):
\begin{equation}
\label{equ:dIbydT}
\Delta T_{\rm CMB}={{\Delta S_{\rm CMB}}\over{S_{\rm CMB}}}{{e^x-1}\over{xe^x}}
T_{\rm CMB},
\end{equation} 
which can also be written as
\begin{equation}
\Delta T_{\rm CMB}={2c^2h^2\over k^3T_{\rm CMB}^2}{\sinh^2(x/2)\over x^4}
 \Delta S_{\rm CMB}.
\end{equation} 
For full accuracy equation~(\ref{equ:dIbydT}) should be
averaged over the bandpass of the filter.  Carrying this out for the 850\mum\
filter (see Holland et al.~1998b)
we find the effective frequency to be $348.4\,$GHz (or 860.5\mum).
This corresponds to ${\overline x}=6.13$ for the 850\mum\ filter,
while the value is 11.9 at 450\mum, which is therefore too far out in the
Wien tail to have any sensitivity to CMB fluctuations.
The resulting conversion factors between Jansky and Kelvin, for the 450 and
850\mum\ data are given in Table~2.

%
%
\begin{table*}
\label{tab:convert}
\begin{center}
\caption{ \hfil Conversion factors between thermodynamic temperature, flux
density and voltage.\hfil }
\begin{tabular}{cllrl}
\noalign{\medskip}
\noalign{\smallskip}
Wavelength (\mum) & Convert from & to & Multiply by & Comment\cr
\hline
450 & Voltage   & Flux density & 800\,Jy/V & \cr
450 & Flux density & Temperature & 49{,}800\,\muk/mJy &
  FWHM=7.5 arcsec \cr
850 & Voltage   & Flux density & 240\,Jy/V & \cr
850 & Flux density & Temperature & 568\,\muk/mJy &
  FWHM=14.7 arcsec \cr
\hline
\end{tabular}
\end{center}
\end{table*}
\bigskip

\section{Placing constraints on an underlying astronomical signal}
\label{sec:likely}
\subsection{Binning the data}
Because of sky rotation, the data for each bolometer cannot be simply
averaged across the entire observing run.  Instead, the timestream is divided
into bins that are some reasonable fraction of the beamsize.
To choose the bin-size, let us restrict our attention 
to the bolometers on the outer ring of the array, which will rotate the
most throughout the course of the observations.  Using the positional 
information in the data, we calculate that the central beam in the three-beam
measurement moved by roughly 6 arcsec (a third of a beamwidth) over the 
entire run.  However, the two `off' beams moved this same distance in just 
one scan.  In addition there is a delay between successive scans, so binning 
more than one scans' worth of data would result in a smeared beam. 
Binning the data any finer than one scan would simply result in a more unwieldy
and noisier data-set, with no additional information.
This would not be the case if the bolometer noise varied throughout
a scan, but for our data-set constant noise in each bolometer is a good
approximation.

With these considerations, plus removing those sections of data contaminated 
by atmospheric emission, we obtain 175 binned spatial points to use in the 
likelihood fits (5 spatial points for each of the 35 usable bolometers).
Since there is potentially significant
information contained in the
off-diagonal correlations between sky positions, we wanted to avoid
binning any more coarsely than this.  In addition we also want to retain
the possibility of negative correlations introduced between `on' beams and
`off' beams when we perform our likelihood analysis, as described in the
next section.

\subsection{Likelihood analysis}
Assuming that the noise and underlying astronomical signal are Gaussian
distributed, all the information is contained in the two-point correlation
function,
$\left\langle{\Delta T}(\hat{n}_i){\Delta T}(\hat{n}_j)\right\rangle$,
where $\hat{n}$ 
is a unit vector on the sky, and $i,j$ correspond to two data points.  
The  traditional approach is to take a parameterized theory that describes 
the signal and perform a Bayesian likelihood analysis on the data in order to 
determine the value of the parameters.
We consider here the two simplest models for underlying sky
fluctuations, the flat power spectrum, and
the Gaussian AutoCorrelation Function (GACF).

\subsubsection{The flat power spectrum ($Q_{\rm flat}$)}
\label{sec:qflat}
CMB anisotropies can be described by an expansion of spherical harmonics on the
sky.  The angular power spectrum of the amplitude of the spherical modes
is given by $C_\ell$, and completely describes the temperature correlations
between two spots on the sky:
\begin{eqnarray}
\label{equ:cth_1beam}
C(\theta_{ij}) \equiv & \hspace*{-3.25cm}\left\langle
   \Delta T(n_i) \Delta T(n_j) \right\rangle \nonumber\\
 = & {{1}\over{4\pi}}\sum_{\ell=2}^{\infty}
 (2\ell+1) C_\ell P_\ell(\hat{n_i}\cdot \hat{n_j})\,
 {\rm e}^{-\ell(\ell+1)\sigma^2},
\end{eqnarray} 
where $P_\ell$ are the Legendre polynomials and 
$\cos\theta_{ij}=\hat{n_i}\cdot \hat{n_j}$.  The exponential term at the 
end accounts for the SCUBA Gaussian beam shape, with Gaussian width
given by $\sigma$.

The correlation model is encoded in the $C_\ell$, and is the quantity we
are trying to fit.   For the specific case of a scale-invariant
Sachs--Wolfe, or `flat' power spectrum, this takes the form of 
\begin{equation}
\label{equ:c_l_flat}
C_\ell={{24\pi}\over{5}}{{Q_{\rm flat}^2}\over{\ell(\ell+1)}},
\end{equation}  
where $Q_{\rm flat}$ is the amplitude parameter to fit using the data
(and explicitly is the extrapolation to the quadrupole for a flat spectrum).

For an experiment with $N$ data points, we compute
equation~(\ref{equ:cth_1beam})
for each pair of points and construct $C_{ij}^{\rm th}$, the 
{\em theoretical} correlation matrix.  It completely describes the 
relationships between the data, and depends on the model of the anisotropy 
(the $C_\ell$), the positions on the sky that the experiment studies 
(the Legendre polynomials), and the beam response function (the exponential).  

These last two terms can be considered as a weighting function of the 
power spectrum, which is called the `window function' \cite{whiteandsred} of 
the experiment,
\begin{equation}
\label{equ:wl_simp}
W_\ell(\theta_{ij}) = P_\ell(\hat{n_i}\cdot \hat{n_j})
 {\rm e}^{-\ell(\ell+1)\sigma^2},
\end{equation} 
and can be used to judge the range of scales to which an experiment is
sensitive.

This is the simplified case for an experiment that measures the temperature of
single spots on the sky.  A SCUBA data point actually
consists of three beams with weights $w=[-1,2,-1]$:

\begin{equation}
\label{equ:T_prime}
\Delta T_i^\prime= \sum_{\alpha=1}^3 w_\alpha \Delta T(\hat{n}_i^\alpha),
\end{equation}
which changes the window function and correlation matrix to
\begin{equation}
\label{equ:wl_manybeam}
W_\ell(\theta_{ij}) = \sum_\alpha^3 \sum_\beta^3 w_\alpha w_\beta
P_\ell(\hat{n}_i^\alpha \cdot \hat{n}_j^\beta) {\rm e}^{-\ell(\ell+1)\sigma^2},
\end{equation} 
\begin{equation}
\label{equ:cth_manybeam}
C_{ij}^{th} = \sum_\alpha^3 \sum_\beta^3 w_\alpha w_\beta
 C(\cos^{-1}(\hat{n}_i^\alpha \cdot \hat{n}_j^\beta)).
\end{equation} 

\begin{figure}
\begin{center}
\leavevmode \epsfysize=8cm \epsfbox{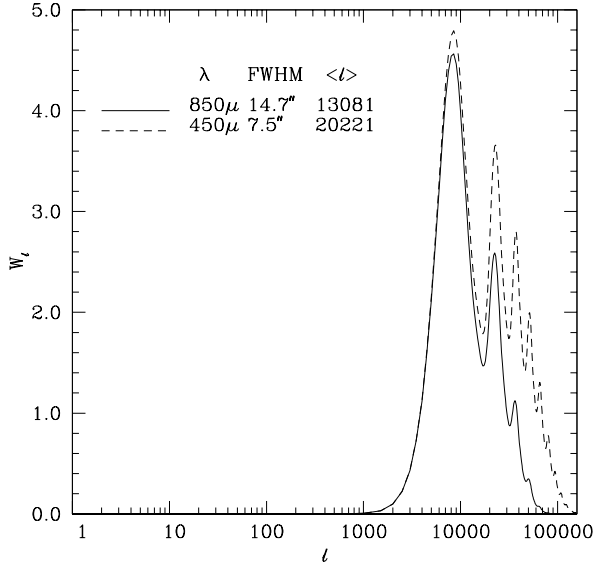}
\end{center}
\caption{The zero lag window functions for the double-difference
SCUBA set-up with a beam throw of 90 arcsec.  The high $\ell$
behaviour is set by the beamsize, while the low $\ell$ shape would be
different for other beam throws.}
\label{fig:wl}
\end{figure}

For estimating the scale to which the experiment is sensitive, the off-axis
contributions ($i\neq j$) are ignored. Then equation~(\ref{equ:wl_manybeam})
reduces to the window function at {\em zero lag}:
\begin{equation}
\label{equ:cth_1beam_zl}
W_\ell = {\rm e}^{-\ell(\ell+1)\sigma^2}[3-4P_l(\cos\gamma)+P_l(\cos2\gamma)],
\end{equation}
where $\gamma$ is the chop amplitude. Fig.~\ref{fig:wl} is a plot of the 
zero lag window functions relevant for a SCUBA chop of 90 arcsec.
The effective center, $\left\langle\ell\right\rangle$
is derived by taking the window function weighted with a flat power spectrum.  

Once we have the correlation matrix, we can fit for the the parameter
$Q_{\rm flat}$ using
\begin{equation}
\label{equ:lq}
L(Q_{\rm flat})\propto{{1}\over{|K|}}
 \exp\left(d\cdot K^{-1}(Q_{\rm flat})\cdot d^T\right),
\end{equation}
\begin{equation}
\label{equ:k}
{\rm with}\quad
K_{ij}(Q_{\rm flat})=C_{ij}^{th}(Q_{\rm flat}) + {\cal{N}}_{ij} 
\end{equation}
\cite{bunnetal,srednickietal}.
Here $d$ is the $1\times N$ vector of measured data, and $\cal N$ is the noise
correlation matrix.  We assume that the data have uncorrelated noise signals
after sky subtraction, and thus ${\cal{N}}_{ij} = \sigma_i\sigma_j\delta_{ij}$.

The matrix inversion can be solved using standard techniques such as singular 
value decomposition, but it is more computationally efficient to use advanced 
methods, of which signal to noise eigenmode decomposition is the most common
(see e.g.~Bond 1995; Bunn~1995; Knox~1997),
when the number of data points is on the order of 200 or so.  

Note that in previous CMB experiments at these scales, it was possible to 
neglect off-axis contributions because of the large angular separation between
data points.  This is not the case for SCUBA, as the pixel separation on
the array is comparable to the beamsize and is only a factor of 4 smaller 
than the chop amplitude for this data-set.

\begin{figure}
\begin{center}
\leavevmode \epsfysize=8cm \epsfbox{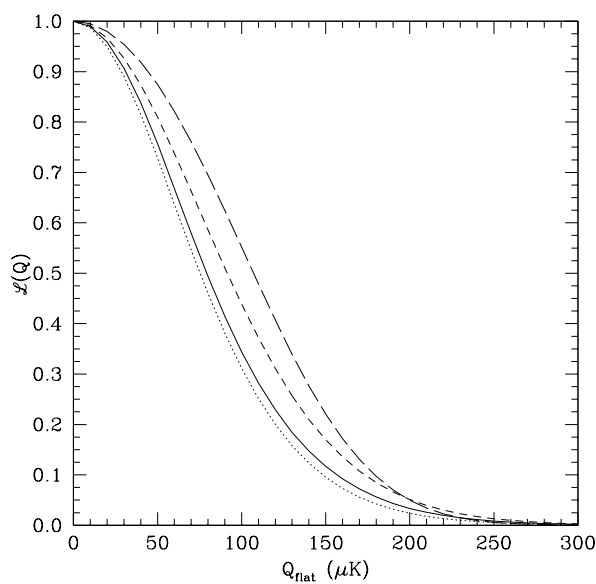}
\end{center}
\caption{The likelihood function for the flat power spectrum model.
The solid line shows our best estimate of the likelihood function, with
subtraction of sky-planes using the 850\mum\ data, and including the full
correlation matrix.
The dotted line is the result obtained using the 850\mum\ mean
for sky subtraction, while the short-dashed line shows the limit using
the 450\mum\ subtraction.  The long dashed line is the 850\mum\ corrected
data had we ignored off-axis contributions to the correlation matrix.}
\label{fig:qflat}
\end{figure}

A plot of the likelihood function over a range of $Q_{\rm flat}$ is given in 
Fig.~\ref{fig:qflat}.  The results indicate that the level of anisotropy is
consistent with zero, with an upper limit of 143\muk\ (95 per cent
confidence).  
This result was obtained using the 850\mum\ mean for sky subtraction, and
is shown by the dotted line in the figure.  
The upper limit when the 450\mum\ mean is used is 164\muk, approximately
7 per cent higher, as expected from the previous discussion, and is
shown by the short-dashed line.
The long-dashed curve in Fig.~\ref{fig:qflat} was generated by removing the
off-axis contributions to the correlation matrix.  This would have given a
substantially different (and incorrect) answer, verifying
the importance of retaining these terms in the analysis.
Note that we have integrated the likelihood using a uniform prior distribution
in $Q_{\rm flat}$.  Different choices of prior would affect the results
somewhat, although the likelihood falls off so rapidly at high $Q_{\rm flat}$
that it is hard to increase the upper limit dramatically by any reasonable
choice of prior.

As a test we removed bolometers 32 and 36 from the data and 
redid the analysis.  As mentioned earlier, these bolometers show stronger 
evidence of correlation than other pairs.  The upper limit falls,
not surprisingly, to 136\muk\ when these signals are removed.
We also tested the effect of removing the noisy bolometers, and found that 
they do not affect the  $Q_{\rm flat}$ limit appreciably. Furthermore, the 
effect of removing a plane from the data, as described in 
Section~\ref{sec:data_reduc} has only a small effect, as we expected from the
small gradients that we found.  Nevertheless, we consider the data-set with the
sky-planes subtracted from the timestream to yield our best estimate of the
likehood function, and so we show this curve with a solid line in
Fig.~\ref{fig:qflat}.  The upper limit is 152 \muk\ (95 per cent confidence).

\subsubsection{Gaussian Auto-Correlation Function (GACF)}
Although largely replaced by the flat power spectrum in most recent analyses
of CMB anisotropy, it is still
useful to calculate the result obtained assuming a GACF as a model for the
correlations.  This is particularly true since at these angular scales, we
do not expect any primary CMB fluctuations, and so any actual signals
present could be of any form, even one with a preferred correlation
scale.  For a GACF assumption,
the theoretical correlation between two beams of Gaussian
width $\sigma$ separated on the sky by an angle $\theta_{ij}$ is given by
\begin{equation}
\label{equ:gacf}
C(\theta_{ij})={{C_0\theta_c^2}\over{2\sigma^2 + \theta_c^2}}\exp
\left[{{-\theta_{ij}^2}\over{2(2\sigma^2 + \theta_c^2)}}\right].
\end{equation} 
Here $C_0$ is the amplitude of the fluctuations, and $\theta_c$ is the sky
coherence angle.

Again, we construct the correlation matrix for a three beam experiment using
equation~(\ref{equ:cth_manybeam}), and redo the likelihood analysis.  In this
case, we need to evaluate the likelihood function over a two dimensional
parameter space in order to fit for $C_0$ and $\theta_c$.  

\begin{figure}
\begin{center}
\leavevmode \epsfysize=8cm \epsfbox{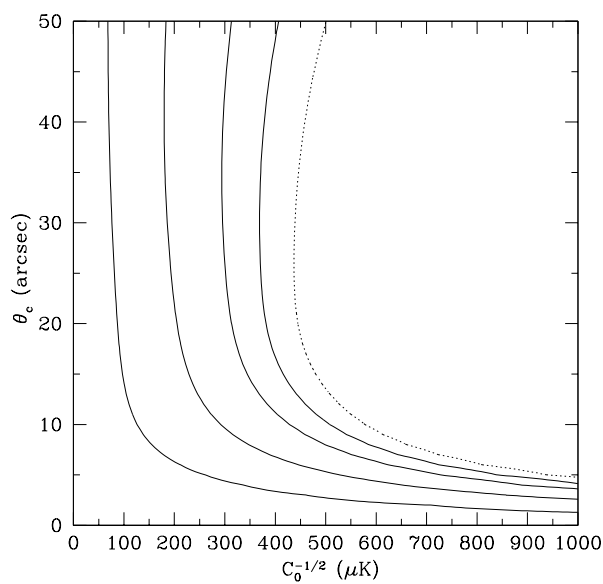}
\end{center}
\caption{The likelihood contours (90, 50, 20 and 10 per cent of the peak)
for the two parameter GACF model are shown as solid lines.
The dotted line shows the 95 per cent two dimensional
confidence region.  The most sensitive coherence length is 23 arcsec,
where $C_0^{1/2}{<}\,$355\muk\ (marginalised 95 per cent confidence limit).}
\label{fig:gacf}
\end{figure}

The result, shown in Fig.~\ref{fig:gacf} for the data sky-subtracted
using the 850\mum\ mean, is $C_0^{1/2}{<}\,$355\muk\ (95 per cent confidence),
at a most sensitive coherence angle of 23 arcsec.  To be explicit, we
integrated using a uniform prior probability distribution in the
quantity $C_0^{1/2}$.
It is customary to use $C_0^{1/2}$
as a measure of the rms temperature sensitivity to compare with other
experiments; We obtain
$\Delta T/T < 1.1\times10^{-4}$, which is a full order of magnitude 
lower than the Church et al.~(1993) result using the precursor
to SCUBA, UKT14,
where the wavelength and coherence angle are are essentially identical.
Also note that $Q_{\rm flat}/C_0^{1/2}\simeq0.4$, and 
$1/\theta_c \simeq \left\langle\ell\right\rangle$,
which are the expected comparisons between these
analysis methods \cite{whiteandscott}.
Fig.~\ref{fig:gacf} makes it clear that the data are not strongly 
correlated, as the confidence limit in the GACF is fairly broad in
coherence angle.  Of course, we expect this result based on our 
$Q_{\rm flat}$ analysis, where the most likely correlation was zero.

\section{Contribution of unresolved point sources to the CMB Power Spectrum}
\label{sec:pointsource}
In the usual notation, $C_\ell$ denotes the power spectrum of CMB anisotropies
on the sky, which is the spherical harmonic analogue of the Fourier
power spectrum on a flat plane.  Just as in the Fourier case, a population
of Poisson-distributed point sources will contribute equally to the power
spectrum on all scales, i.e.~we expect to obtain $C_\ell={\rm constant}$
from random point sources.  The amplitude of this contribution clearly
depends on the source counts of the population, and can be calculated as
\begin{equation}
\label{eq:clsources}
C_\ell(\nu)=\int_0^{S_{\rm cut}} {d{\widetilde N}\over dS_\nu}S_\nu^2\,dS_\nu,
\end{equation} 
(see e.g.~Tegmark \& Efstathiou~1996; Scott \& White in preparation).
Then we convert flux units to
thermodynamic temperature units, as discussed in
Section~\ref{sec:calibration}.  Note, however that here we are
interested in the flux rather than flux per beam, and so we use the same
conversion, {\it without} the beam solid angle.
In equation~(\ref{eq:clsources}) $\nu$ is the particular frequency under
consideration, $S_\nu$ is the flux, $d{\widetilde N}/dS_\nu$ is the
differential source count (i.e.~${\widetilde N}(S_\nu)$ is the number of sources
per steradian {\it fainter} than $S_\nu$), and $S_{\rm cut}$ is some flux above
which we feel confident we can identify individual sources.  All reasonable
source counts give convergent $C_\ell$ at the faint end,
and generally the precise position of the upper flux cut will not be critical.
Integrating by parts, and switching to the more conventional notation
$N\equiv N({>}S_\nu)$, the number density {\it brighter} than some flux
threshold, we obtain
\begin{equation}
C_\ell(\nu)=2\int_0^{S_{\rm cut}}\!N\,S_\nu\,dS_\nu
 - N(S_{\rm cut})\,S_{\rm cut}^2.
\end{equation} 

In order to estimate what we might expect, we have performed this integral
using a phenomenological fit to the current SCUBA counts.  We find that
a simple two power-law fit of the form
\begin{equation}
N=A \left({S_\nu\over S_0}\right)^{-\alpha}
  \left(1+{S_\nu\over S_0}\right)^{-\beta}
\end{equation} 
fits all the available data (Smail, Ivison \& Blain~1997; Barger et al.~1998;
Holland et al.~1998a; Hughes et al.~1998, Lilly et al.~in preparation),
and has the same shape as some successful
models for star formation history (e.g.~fig.~11.(b) of Blain et al.~1998),
with $A\simeq{\rm few}\times10^6\,{\rm Sr}^{-1}$, $S_0\simeq10\,$mJy,
$\alpha\simeq0.8$ and $\beta\simeq1.8$.  For definiteness we normalise to
the \cite{Hugetal} counts from the Hubble Deep Field (HDF), since this seems
to be currently the most robust.  However, the small
numbers of sources found in {\it any} of the deep surveys allows a wide
range of possible normalizations.  A total of 5 sources were detected in
the HDF above $3\,$mJy, and so the 90 per cent confidence range for Poisson
statistics spans the range 2.0--10.5.  We use the central value of
$A=4.1\times10^6\,{\rm Sr}^{-1}$ to give a specific model for estimating
the expectations, keeping in mind that a factor of 3 lower or 2 higher would
not be very unlikely.  Although our parameterized model is a reasonable one,
we wish to stress that it is just a
convenient example; In reality the shape is only constrained over a narrow
range of fluxes, and of course could be quite different at either the bright
or faint ends.
\begin{figure}
\begin{center}
\leavevmode \epsfysize=8cm \epsfbox{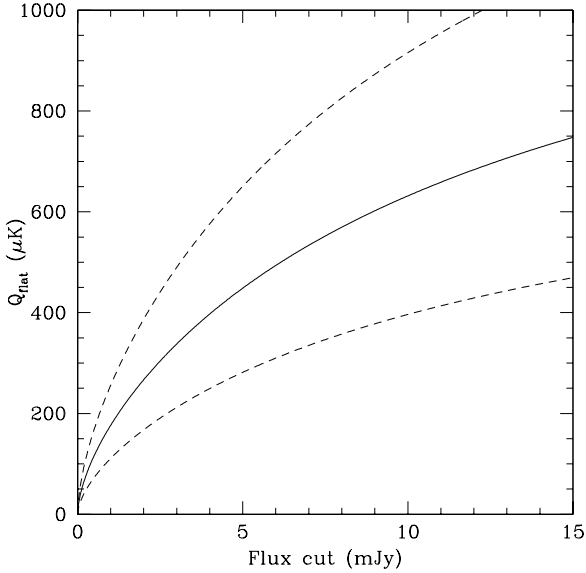}
\end{center}
\caption{$Q_{\rm flat}$ estimates based on the unresolved source count model
described in the text.  The solid line is the normalization based on the 5
sources detected in the Hubble Deep Field by Hughes et al.~(1998), with the
90 per cent confidence region enclosed by the dashed lines.}
\label{fig:qflat_sources}
\end{figure}

We plot the resulting power spectrum in Fig.~\ref{fig:qflat_sources} as a
function of the flux cut for bright sources.  Explicitly, we have
converted to the equivalent amplitude of a flat power spectrum through
the experimental window function (see Section~\ref{sec:qflat}),
as measured by the
expectation value of the quadrupole $Q_{\rm flat}$.  The dashed lines on the
plot represent the 90 per cent CL range from the HDF normalization.
We see that an equivalent $Q_{\rm flat}$ of higher than around $200\,\mu$K
would be expected for even the most drastic cut of $2\,$mJy.  Since this is
close to the rms in the data-set that we analysed, our expectation would
be that we could set a limit no lower than about $200\,\mu$K.

For sources below the break, $S\ll S_0$ the integral becomes analytic and
\begin{equation}
C_\ell = {A\alpha\over 2-\alpha}\left({S_0\over S_{\rm cut}}\right)^\alpha
 S_{\rm cut}^2,
\end{equation} 
which is just $(\alpha/(2-\alpha))N S_{\rm cut}^2$.  Hence for a particular
normalization of the counts, and for small values of the slope,
the value of $Q_{\rm flat}$ will vary approximately as
$\sqrt{\alpha}$.  So analysis of fluctuations in SCUBA data-sets, such as
the one presented here, are sensitive to the slope of the counts at the
faint end.

The phenomenological model gives a central value of 267\muk,
even for a flux cut as low as
$2\,$mJy.  We note that if we were to normalise the counts to
effectively higher values, such as the Smail et al.~(1997)
central value, we would
predict fluctuations about 1.3 times higher.
Recognising that we should have seen evidence of fluctuations in the data
if there were a large number of faint sources, we can use our limit to
constrain the faint end slope.  To do this we fix the flux cut to be
$2\,$mJy, which is very conservative for this data-set, and also
convenient since it corresponds to the flux cut for the HDF counts
\cite{Hugetal}.
Then we can fold our likelihood distribution for $Q_{\rm flat}$
together with the Poisson probability distribution for the 5 HDF sources
to calculate a likelihood function for
the faint end slope.  We find a 95 per cent confidence limit of
$\alpha<0.52$, which is already shallower than some of the models.
We stress that this is quite conservative in terms of assuming the lowest
feasible flux cut for our data-set.
More stringent constraints could possibly be placed by calculating
counts and fluctuations for detailed models, but this is probably
not warranted for the current modest data-set.

The above discussion considered only sources that are distributed
randomly on the sky.
On large scales it is reasonable to assume that the clustering of distant
sources does have a negligible effect.  This is because any three dimensional
clustering will be washed out when projected in two dimensions.  However,
this is no longer justifiable at the smallest angular scales probed, and
particularly for the sources which may dominate fluctuations detectable by
SCUBA, namely dusty star-forming galaxies at high redshift.  In hierarchical
clustering scenarios we expect galaxies and clusters to have formed from
the build up of smaller units at earlier times.  The angular scales probed
by SCUBA lie in the range which is typical for a rich cluster of galaxies
at high redshift.  Hence we might expect to see dusty galaxies clustering
together as they form a rich cluster, or sub-clumps of large galaxies being
assembled on group scales, or perhaps galaxies flaring up in star formation
as they fall into groups or are gobbled up by centrally dominating cluster
galaxies.  It is to be hoped that issues such as these may be addressed
using detailed fluctuation analyses of the deepest SCUBA integrations.

\section{Conclusions}
\label{sec:conclude}
The data-set analysed here was
obtained using less than an hour of integration time, 
which makes it significantly shorter than most other CMB experiments!
In order to improve on these results, a deeper integration is necessary.
The average signal-to-noise for the binned data points in the likelihood
fits is roughly 0.1, although we erred on the side of under-binning.
Most CMB experiments aim for a S/N of unity, which is generally
considered to be the optimal compromise between sky coverage
and integration time.  This is certainly feasible with more ambitious
SCUBA integrations, including existing data-sets.  It should be pointed
out however, that there may be additional challenges involved with
applying correlation analyses to data taken in the mapping mode.

Using the 450\mum\ channel as an atmospheric monitor does introduce
more noise in the final results, but at levels of only a few percent.  This
indicates that another method of sky removal is possible, if there is
good reason not to use the 850\mum\ data-set itself.
There is also some evidence of residual atmospheric gradients across the SCUBA 
array, but they are small, and do not appreciably change our results.  
With the growing amount of blank field SCUBA data, both these effects can and
should be studied in more detail.

We have presented here the best upper limit of CMB anisotropy on arcsecond
scales at 850\mum.  A careful treatment of the data was performed in order
to test for non-astronomical signals.  Our best estimate for the likelihood
function yields an upper limit on the flat-spectrum-extrapolated quadrupole
of $Q_{\rm flat}<152$\muk.  Using a recent estimate of the source counts at
$2\,$mJy we can convert this into an upper limit on the slope of the faint
counts, since otherwise we would have detected the fluctuations due to even
Poisson-distributed sources.  We find that if $N(>S)\propto S^{-\alpha}$
then $\alpha<0.52$ at the 95 per cent confidence limit.

Larger SCUBA data-sets already exist, and these could easily be subjected
to similar analysis techniques to those discussed here.
We fully expect that future data-sets will yield detections, since the
fluctuations expected in most favoured models lie only just below what
we could have detected here.  Investigation of such signals should provide
independent
constraints on the population of sources which are just below the detection
threshold, and should also yield new insight into the clustering properties
of the SCUBA-bright objects.  It is also of course possible that other
physical effects will be important at these wavelengths and angular scales,
for example Sunyaev-Zel'dovich effects, or perhaps currently unconsidered
foreground processes within our own Galaxy.  In any case, fluctuation
studies with SCUBA are likely to be increasingly important for understanding
the sub-millimetre and far-infrared sky, and in particular for planning
future satellite missions such as FIRST and Planck.

\section*{ACKNOWLEDGMENTS}
We wish to thank Henry Matthews for help with the JCMT observations,
and Martin White and Greg Fahlman for helpful comments. CB
wishes to thank Tim Jenness for his valued assistance in extracting the data
and Wayne Holland for supplying the 850\mum\ filter function.
This work was supported by the National Sciences and Engineering Research
Council of Canada. The James Clerk Maxwell Telescope is operated by The Joint
Astronomy Centre on behalf of the Particle Physics and Astronomy Research
Council of the United Kingdom, the Netherlands Organisation for Scientific
Research, and the National Research Council of Canada.

\bsp

\end{document}